\documentclass[a4paper,twocolumn,twoside]{article}
\newcommand{\affil}[1]{$^{\rm #1}$}
\usepackage{epsfig}
\usepackage[authoryear]{natbib}

\bibpunct{(}{)}{;}{a}{}{,}

\usepackage{graphicx}
\date{} 
\def\spose#1{\hbox to 0pt{#1\hss}}
\def\lta{\mathrel{\spose{\lower 3pt\hbox{$\mathchar"218$}}
     \raise 2.0pt\hbox{$\mathchar"13C$}}}
\def\gta{\mathrel{\spose{\lower 3pt\hbox{$\mathchar"218$}}
     \raise 2.0pt\hbox{$\mathchar"13E$}}}

\def\aj{{AJ}}%
\def\araa{{ARA\&A}}%
\def\apj{{ApJ}}%
\def\apjl{{ApJ}}%
\def\aap{{A\&A}}%
\def\aapr{{A\&A~Rev.}}%
\def\baas{{BAAS}}%
\def\mnras{{MNRAS}}%
\def\pasp{{PASP}}%
\def\nat{{Nature}}%
\def\planss{{Planet.~Space~Sci.}}%
\title{\large\bf\flushleft
On the Possibility of Habitable Moons in the System of HD~23079: \\
Results from Orbital Stability Studies}

\author{\parbox{\textwidth}{\flushleft
\vspace{-0.5cm}
{\it M. Cuntz\affil{A,B}, B. Quarles\affil{A,C}, J. Eberle\affil{A,D}, and A. Shukayr\affil{A}}\\
\vspace{0.4cm}
{\small \affil{A}\,Department of Physics, University of Texas at Arlington, Arlington,
                   TX 76019, USA} \\
{\small \affil{B}\,Corresponding author. Email: cuntz@uta.edu} \\
{\small \affil{C}\,Present address: Space Science and Astrobiology Division 245-3,
                   NASA Ames Research Center, Moffett Field, CA 94035, USA} \\
{\small \affil{D}\,Present address: North Lake College, Central Campus, Irving, TX 75038, USA}
}}
\begin{document}
\twocolumn[
\begin{changemargin}{.8cm}{.5cm}
\begin{minipage}{.9\textwidth}
\vspace{-1cm}
\maketitle
%
%
\small{\bf Abstract:}
The aim of our study is to investigate the possibility of
habitable moons orbiting the giant planet HD~23079b, a
Jupiter-mass planet, which follows a low-eccentricity orbit
in the outer region of HD~23079's habitable zone.  We show
that HD~23079b is able to host habitable moons in prograde
and retrograde orbits, as expected, noting that the outer stability limit for
retrograde orbits is increased by nearly 90\% compared to that
of prograde orbits, a result consistent with previous generalized
studies.  For the targeted parameter space it was found
that the outer stability limit for habitable moons varies between
0.05236 and 0.06955~AU (prograde orbits) and between 0.1023 and 0.1190~AU
(retrograde orbits) depending on the orbital parameters of
the Jupiter-type planet if a minimum mass is assumed.
These intervals correspond to 0.306 and 0.345 (prograde orbits)
and 0.583 and 0.611 (retrograde orbits) of the planet's Hill radius.
Larger stability limits are obtained if an increased value for
the planetary mass $m_p$ is considered; they are consistent with
the theoretically deduced relationship of $m_p^{1/3}$.  Finally,
we compare our results to the statistical formulae of Domingos et al.
(2006), indicating both concurrence and limitations.  

\medskip{\bf Keywords:}
          astrobiology --- instabilities --- planetary systems ---
          stars: individual (HD~23079) --- stars: late-type



\medskip
\medskip
\end{minipage}
\end{changemargin}
]
\small


\section{Introduction}

\bigskip
\noindent
The existence of planets in orbit about solar-type stars is
now a well-established observational result.  Obviously, the
ultimate quest of this type of study is to discover Earth-type
planets and moons located in the habitable zones (HZs) of their
host stars.  Despite a lack of discoveries, the existence of
Earth-mass planets, including those hosted by solar-type stars,
is strongly implied by various observational findings including
the occurrence and mass distribution of close-in super-Earths,
Neptunes, and Jupiters \citep{how10}.  Measurements by \cite{how10}
indicate an increasing planet occurrence with decreasing planetary
mass $m_p$ akin to $m_p^{-0.48}$, implying that 23\% of stars harbor
a close-in Earth-mass planet (ranging from 0.5 to 2.0 $M_\oplus$);
see also \cite{mar00} for earlier results.  More recent results
were given by \cite{wit11}; they indicate based on the analysis
of 67 solar-type stars that nearly 20\% of stars are expected
to host planets with a minimum mass of less than 10 $M_\oplus$.

There is also a range of studies focusing on exosolar moons,
especially moons possibly located in stellar HZs.  Previous
studies on exosolar moons include work by \cite{wil97},
\cite{bar02}, and more recently by, e.g., \cite{kip09},
\cite{kal10}, and \cite{wei10}.  \cite{wil97} investigated
appropriate orbital parameters of possible moons, and argued
that the moons need to be large enough (i.e., $> 0.12~M_\oplus$)
to retain a substantial and long-lived atmosphere and would also
need to possess a significant magnetic field to be continuously
habitable.  The work by \cite{kip09,kal10,wei10} focused on the properties and detectability of moons
of transiting planets, particularly those which can be detected
as part of the Kepler mission.  \cite{kip09} pointed out that
HZ exomoons down to $0.2~M_\oplus$ may be detected and approximately
25000 stars could be surveyed for HZ exomoons with Kepler's field
of view.  \cite{kal10} conveyed that transmission spectroscopy of
moons detected as part of the Kepler mission has a unique potential
to screen them for habitability in the near future.  In addition,
\cite{wei10} explored the limits on the orbits and masses of moons
around currently known transiting exoplanets, particularly those
discovered by the COROT mission.  Recently, \cite{kip13} showed
seven possible candidates for transiting exomoons using observations
from the Kepler mission.

Additional recent work with implications on the possibility of
habitable exomoons were given by \cite{tin11}, \cite{rob12}, and
\cite{hel13}.  \cite{tin11} focused on the detection of a gas-giant
exoplanet around HD~38283, a G0/G1~V star, in an eccentric orbit with
a period of almost exactly one year; a detailed discussion about the
potential existence of exomoons and exo-Trojans (see below) is conveyed
by the authors.  Another interesting study about gaseous planets
about solar-type stars permitting the principal possibility of
habitable exomoons was given by \cite{rob12}.  \cite{hel13} provide
a thorough analysis of exomoon habitability constrained by stellar
illumination and tidal heating.

Moon formation scenarios based on a Solar System-like nebula
have shown that a mass scaling ratio,
which is $\sim 10^{-4}$, exists preventing gaseous planets using
core-accretion to gain enough material for forming Earth-mass moons
\citep{can06}.  However, this has not dissuaded the possibility of
gaseous planets from acquiring large Earth-mass moons.
Previous studies have indicated that there can be large stability
zones for irregular or captured satellites.  In the Solar System,
Kuiper belt objects \citep[see, e.g.,][]{luu97} have demonstrated
retrograde motion resulting from formation scenarios with
proper conditions \citep{sch08}.  Moreover, many irregular moons
have been discovered where collisions have accreted
to substantial sizes \citep{nes03}.  Currently, the Hunt for
Exomoons with Kepler (HEK) \citep{kip12} has utilized the
hypothesis of acquiring Earth-mass exomoons via capture which
would most likely be identified as orbiting in retrograde.

Long-term orbital stability in stellar HZs is a necessary condition
for the evolution of any form of life, particularly intelligent life.
Examples of previous studies on the orbital stability of hypothetical
Earth-mass planets in stellar HZs include the work by \cite{geh96},
\cite{nob02}, \cite{men03}, \cite{cun03}, \cite{asg04},
\cite{jon05,jon06,jon10}, \cite{kop10} and \cite{ebe11}.  These studies
explore the dynamics of multi-body systems noting that many extrasolar
planetary systems contain more than one extrasolar giant planet (EGP),
which is of pivotal importance for the orbital stability of possible
Earth-mass object(s) in those systems.

For example, in the HD~23079 system,
a Jupiter-mass planet is found to orbit the star
in the HZ, therefore jeopardizing the existence of habitable
terrestrial planets.  This system has previously been studied
by \cite{dvo04}, \cite{sch07}, and \cite{ebe11}.
The study of HD~23079 is of particular interest because (1) the
mass and spectral type of the center star is fairly similar
to the Sun, (2) the Jupiter-mass planet is found to orbit
the star in a nearly circular orbit, and (3) any exomoon of
that planet, if such a moon exists, would therefore exhibit
a relatively stable climate.

In systems like HD~23079, there is the general possibility
of habitable Trojan planets as well as habitable exomoons.
A Trojan planet is one that librates around
one of the Lagrangian points L4 and L5 of the giant planet.
These points lie on the giant planet's orbit, ahead (L4) and
behind (L5) the planet, each forming an equilateral triangle
with the planet and its star.  \cite{dvo04} investigated the
stability regions of hypothetical terrestrial Trojans librating
around L4 and L5 in specific systems, including HD~23079, and
obtained relationships between the size of the stability regions
and the orbital parameters of the giant planets, particularly its
eccentricity.  \cite{sch07} thereafter identified numerous
exoplanetary systems that can harbor Trojan planets in stable
orbits.  All these systems also provide the opportunity for
hosting exomoons around Jupiter-type planets, including
exomoons located in stellar HZs.  Updated results
for HD~23079 given by \cite{ebe11} showed that this system is
also well suited for the existence of habitable Earth-type Trojans,
especially if a generalized stellar HZ is assumed,
which in the Solar System corresponds to 0.95 to 1.67~AU from
the Sun \citep{kas93}; see Sect. 2.2 for details.

A more recent example of a system with the potential of hosting
both exomoons and exo-Trojans in the ``habitability phase space"
is HD~38283 as described by \cite{tin11}.  This system is known
to possess an EGP around its host star largely or entirely located
in the stellar HZ considering that HD~38283b's orbit
is highly eccentric (i.e., $e_p \simeq 0.41$), a notable difference
to HD~23079b considered in the present study.  Exomoons and
exo-Trojans as companions of gas-giant exoplanets in 0.5--2 AU
orbits (somewhat depending on the stellar spectral type) thus
offer the possibility of providing circumstellar habitability.
Exomoons and/or exo-Trojans may be able to form in-situ or be
captured; regarding the latter, examples include the outer
satellites of Jupiter and Saturn \cite[e.g.,][]{jew07}.

In the present study, we focus on the system of HD~23079 with
a concentration on the principal possibility of habitable
exomoons.  Our paper is structured as follows.  First,
we discuss the stellar and planetary parameters of the
HD~23079 system.  Next, we comment on the circumstellar
habitability of HD~23079.  Thereafter, we present our results
and discussion, including a detailed statistical analysis of the
exomoon orbital stability limits as well as a comparison of our
results with the formulae of \cite{dom06}.  Finally, we convey
our summary and conclusions.


\section{Theoretical Approach}

\subsection{Stellar and Planetary Parameters}

HD~23079 has been monitored as part of the Anglo-Australian Planet
Search (AAPS) program \citep{tin02} that is able to perform extrasolar
planet detection and measurements with a long-term, systematic radial
velocity precision of 3 m~s$^{-1}$ or better.  HD~23079 was identified
to host a Jupiter-mass planet in a relatively large and nearly circular
orbit.  HD~23079 is an inactive main-sequence star; \cite{gra06}
classified it as F9.5~V (see Table~1; all parameters have their usual
meaning), an updated result compared to \cite{hou75} who found that
HD~23079 is intermediate between an F8 and G0 star.  Its stellar
spectral type corresponds to a mass of $M = 1.10 \pm 0.15$ $M_\odot$.
The stellar effective temperature and radius are given as
$T_{\rm eff} = 6030 \pm 52$ K and $R = 1.106 \pm 0.022$ $R_\odot$,
respectively \citep{rib03}.  Thus, HD~23079 is fairly similar to
the Sun, though slightly hotter and slightly more massive.
The detected planet (HD~23079b) has a minimum mass of
$m_p \sin i = 2.45 \pm 0.21$ $M_J$.  Furthermore, it has a
semimajor axis of $a_p = 1.596 \pm 0.093$ AU and an eccentricity
of $e_p = 0.102 \pm 0.031$ \citep{but06}, corresponding to an
orbital period of $P = 730.6 \pm 5.7$ days.  The original results by
\cite{tin02} indicated very similar planetary parameters.

The orbital parameters of HD~23079b are relatively similar to those
of Mars, implying that HD~23079b is orbiting its host star in or
near the outskirts of the stellar HZ; see discussion below.  The
existence of HD~23079b, a planet about two and a half times more
massive than Jupiter, makes it difficult for a terrestrial planet
to orbit HD~23079 at a similar distance without being heavily affected
by the giant planet; see results from previous case studies by \cite{nob02}
and \cite{yea11} who focused on the dynamics of HD~20782, HD~188015,
and HD~210277, which are systems of similar dynamical settings.
Concerning HD~23079, previous investigations pertaining to habitable
terrestrial Trojan planets were given by \cite{dvo04} and \cite{ebe11}.


\subsection{Stellar Habitable Zone}

The extent of the HZ of HD~23079 can be calculated
following the formalism by \cite{und03} based on the
work by \cite{kas93}, while also adopting an appropriate
correction for the solar effective temperature used in
the definition of solar luminosity, $L_\odot$.
\cite{und03} supplied a polynomial fit that
allows to compute the extent of the conservative and the
generalized HZ based on prescribed values of the stellar
luminosity and the stellar effective temperature (see Table~2).
Noting that HD~23079 is considerably more luminous than
the Sun, it is expected that its HZ is more extended than
the solar HZ, for which the inner and outer limit of the
generalized HZ were obtained as 0.84 and 1.67~AU, respectively
\citep{kas93}.

The generalized HZ is defined as bordered by the runaway
greenhouse effect (inner limit), where water vapor enhances
the greenhouse effect thus leading to runaway surface warming,
and by the maximum greenhouse effect (outer limit), where a
surface temperature of 273~K can still be maintained by a
cloud-free CO$_2$ atmosphere.  The inner limit of the
conservative HZ is defined by the onset of water loss, i.e.,
the atmosphere is warm enough to allow for a wet stratosphere
from where water is gradually lost by photodissociation and
subsequent hydrogen loss to space.  Furthermore, the outer
limit of the conservative HZ is defined by the first CO$_2$
condensation attained by the onset of formation of CO$_2$
clouds at a temperature of 273~K.

For HD~23079, the limits of the conservative HZ are given
as 1.1378 and 1.6362~AU, whereas the limits of generalized HZ
are given as 0.989 and 1.966~AU (see Fig. 1).  The limits
of the generalized HZ are those employed in our numerical
planetary studies\footnote{The physical limits
of habitability are much less stringent than implied by the
numerical precision of these values; nevertheless, these values
were used for checking if the Earth-mass planet has left
the stellar HZ.}.  The underlying definition of habitability
is based on the assumption that liquid surface water is a
prerequisite for life, a key concept that is also the basis
of ongoing and future searches for extrasolar habitable planets
\citep[e.g.,][]{cat06,coc09}.  The numerical evaluation
of these limits is based on an Earth-type planet with a
CO$_2$/H$_2$O/N$_2$ atmosphere; see work by \cite{kas93}
and subsequent studies.

We point out that concerning the outer edge of habitability,
even less conservative limits have been proposed in the meantime
\citep[e.g.,][]{for97,mis00}.  They are based on the assumption of
relatively thick planetary CO$_2$ atmospheres and invoke strong
backwarming that may further be enhanced by the presence of CO$_2$
crystals and clouds.  However, as these limits, which can be as
large as 2.4~AU in case of the Sun, depend on distinct properties
of the planetary atmosphere, they are not relevant for our study.
Nevertheless, we convey this type of limit for the sake of curiosity
(see Fig. 1), noting that it has properly been adjusted to 2.75~AU
in consideration of the radiative conditions of the planetary host star,
HD~23079.  Moreover, the significance of this extreme limit
has meanwhile been challenged based on detailed radiative transfer
simulations \citep{hal09}.

A more thorough investigation of exoplanet and exomoon habitability
points to the consideration of additional factors that might
influence the facilitation of an habitable environment.  Detailed
reviews by, e.g., \cite{lam09} and \cite{hor10} indicate the
importance of magnetic fields and plate tectonics as well as
the prospects for the object of being hydrated as potentially
decisive.  Exomoons of HD~23079b are expected to be in a fortunate
situation from that point of view as one would expect that tidal
effects of the EGP on the moon might result in a hot interior,
and hence make plate tectonics and a magnetic field more likely.
Moreover, if the giant planet had migrated in from beyond the
snow line \citep[e.g.,][]{arm07}, it could help hydrate the moon.

As part of our evaluation of habitability around HD~23079, we
explored whether or not the orbit of HD~23079b is indeed fully
located in the HZ of HD~23079.  Thus, we calculated the limits
of the generalized HZ by also considering the uncertainties in
the stellar luminosity and effective temperature while employing
the formalism by \cite{und03} based on second-order polynomials.
The stellar luminosity and effective temperature were assumed
as $1.45 \pm 0.08$~$L_\odot$ and $6030 \pm 52$~K, respectively.
Therefore, the outer limit of the generalized HZ, denoted as
HZ-o (gen.), see Table~2, is obtained as $1.966 \pm 0.054$~AU.

In addition, we calculated the periastron and apastron of HD~23079b,
denoted as $a_{\rm per}$ and $a_{\rm ap}$, respectively, while
also considering the statistical uncertainties in the planetary
orbital parameters, i.e., the semimajor axis and eccentricity,
given as $1.596 \pm 0.093$~AU and $0.102 \pm 0.031$, respectively
(see Table~1).  Hence, the values for the periastron and apastron
are identified as $1.432 \pm 0.094$~AU and $1.758 \pm 0.113$,
respectively (see Table~2).  Our analysis shows that the orbit of
the EGP is almost certainly completely positioned in HD~23079's HZ;
the probability for HD~23079b's orbit to be partially outside the
generalized HZ is only 5\%.  It should be also noted that following
the study by \cite{wil02} brief excursions from the stellar HZ
do not necessarily nullify planetary habitability because the
latter is expected to mainly depend on the average stellar flux
received over an entire orbit.


\section{Results and Discussion}

\subsection{Orbital Stability Simulations}

For our simulations of the HD~23079 system, we consider both the
observed giant planet HD~23079b and an object of one Earth-mass,
i.e., $3.005 \times 10^{-6}$~$M_\odot$, which based on our
choice of initial conditions is expected to constitute a moon
in regard to the giant planet.  Consequently, we execute a grid
of model simulations by considering different starting conditions
for the moon candidate.  The method of integration assumes a
fourth-order Runge-Kutta integration scheme \citep{gar00} and an
eighth-order Gragg-Bulirsch-Stoer (GBS) integration scheme \citep{gra96}.
The code has been extensively tested against known analytical solutions,
including the two-body and restricted three-body problem \citep[see][for
detailed results]{nob02,cun07,ebe08}.

We limit our computations to a simulation time of $10^6$~yrs and
apply a time-step of $10^{-4}$~yrs for the integration scheme;
the latter is found to be fully appropriate.
In conjunction with previous investigations of habitable Trojan planets
in the HD~23079 system \citep{ebe11}, we pursued test simulations comparing
the planetary orbits based on three different integration time-steps,
which were: $10^{-3}$, $10^{-4}$ and $10^{-5}$~yrs.  In particular, we
evaluated $\Delta R_{ij}$, i.e., the magnitude of the difference between
the position of the planet when different step sizes of $10^{-i}$ and
$10^{-j}$ were used.  We found no discernible differences in the outcome
between models with time-steps of $10^{-4}$ and $10^{-5}$~yrs, thus
justifying the usage of $10^{-4}$ as time-step for this type of simulations.

The giant planet always starts out at the 3 o'clock position relative
to the star placed at the origin, which is assumed to coincide with its
periastron position.  The initial condition (i.e., starting velocity) for
the orbit of the terrestrial moon is chosen such that it is assumed to
be coplanar regarding the motions of the star and the EGP, as well as
in a circular orbit about the EGP, although it is evident that
it will be significantly affected immediately by gravitational pull of
the star, which is expected to prevent from continuing its circular motion.
Furthermore, the moon always starts at the 9 o'clock position relative
to the giant planet.  Our set of models is defined by five values for
the semimajor axis $a_p$ and five values for the eccentricity $e_p$
of the giant planet, which range from 1.503 to 1.689~AU in increments
of 0.0465  and from 0.071 to 0.133 in increments of 0.0155, respectively.
For the EGP, we assume its minimum mass value of 2.45 $M_J$ as default.
However, for models with $a_p = 1.596$~AU and $e_p = 0.102$, the mass of
the EGP is increased by factors of 1.5 and 2.0.


\subsection{Derivation of Orbital Stability Limits}

In our model simulations we consider both prograde and retrograde
orbits of the moon candidate.  In the selected coordinate system
the giant planet orbits the star in a counter-clockwise sense,
which means that for prograde orbits, the moon orbits the giant
planet in a counter-clockwise sense as well.  Hence, for retrograde
orbits, the moon orbits the giant planet in a clockwise sense.
The overwhelming majority of our simulations assumes a minimum
mass for the giant planet; these models will be the focus of the
following methodology.

The identification of model-dependent orbital stability limits
for the moon requires a large number of simulations since the
stability limits (i.e., maximum radii of possible orbits)
can only be determined through trial-and-error.
As the selected approach, both for prograde and retrograde orbits,
a two-step process is adopted.  First, we consider a preliminary
$3 \times 3$ grid for the targeted array of values for the
semimajor axis $a_p$ and eccentricity $e_p$ of the EGP;  for $a_p$,
they are given as 1.503, 1.596, and 1.689~AU, whereas for $e_p$,
they are given as 0.071, 0.102, and 0.133.

For each grid value, a simulation time of $5 \times 10^4$~yrs
is used to obtain estimates for the stability limit
of each ($a_p$, $e_p$) combination.  For each prograde model,
we perform a total of 201 simulations, based on intervals
for the moon's starting distance with 0.0520 and 0.0720~AU
as lower and upper limits, respectively.  Thus, the simulations
are advanced in increments of $1 \times 10^{-4}$~AU regarding
the moons starting distance.  For each retrograde model, we simulate
a total of 201 initial conditions as well; here the limits of the
intervals for the moon's starting distance are given as 0.1050
and 0.1250~AU.  This approach amounts to a total of nearly 3600
simulations for the prograde and retrograde models combined.

To refine our search for orbital stability limits, a second
step is implemented involving two aspects, which are: augmenting
the preliminary coarse $3 \times 3$ grid to the full $5 \times 5$
grid, while also considering substantially longer simulation
times for each model, i.e., target timespans of $10^6$ years.
For each ($a_p$, $e_p$) combination, these long-term simulations
again utilize intervals for setting the moon's starting distance.
For each grid combination, we execute a total of 16 simulations
centered at the expected value for the model-dependent
orbital stability limit.

For models already part of the preliminary $3 \times 3$ grid,
the expected values for the orbital stability limits are obviously
given by the results of the previous short-term computations.
For models beyond the preliminary $3 \times 3$ grid, the expected
values for models with intermediate values of semimajor axis (i.e.,
1.5495 and 1.6425~AU) are generated through applying a polynomial
fitting assuming constant values for $a_p$ (i.e., 1.503, 1.596,
and 1.689~AU).  The polynomial fitting process also assisted us
in estimating where stability limits should occur for intermediate
values of eccentricity (i.e., 0.0865 and 0.1175).

Our results are given in Tables 3, 4, and 5; for convenience,
the data of Table 3 and 4 are also depicted in Figs.~2 and 3. 
Based on the total number of simulations, it is found that,
if the EGP is assumed to possess minimum mass, the stability
limit for habitable moons varies between 0.05236 and 0.06955~AU
(prograde orbits) and 0.1023 and 0.1190~AU (retrograde orbits)
depending on the orbital parameters of the giant planet.  By
keeping the EGP's mass and the semimajor axis fixed, expressions
are encountered structurally similar to those of \cite{hol99};
however, with different coefficients naturally related to our
particular study.

Somewhat larger stability limits of the candidate moon are
obtained if an increased value for the mass of the EGP is
considered.  Specifically, if the mass of the giant planet
is increased by factors of 1.5 and 2.0, the stability limit
for prograde orbits increases from 0.05712 to 0.06290 and
0.06940~AU, respectively.  Furthermore, for retrograde orbits,
the stability limit of habitable moons is found to increase
from 0.1109 to 0.1291 and 0.1400~AU, respectively.


\subsection{Examples}

Examples of orbital stability and instability of exomoons around
HD~23079b are given in Figs. 4 to 8.  The orbital parameters of
the giant planet considered in these case studies are
$a_p = 1.596$~AU and $e_p = 0.102$; additionally, the EGP
is assumed to be of minimum mass.  Figures 4 and 5 indicate that
for prograde orbits short-term ($\simeq 10^4$ years) stability ---
subsequently confirmed to also constitute long-term stability;
see Sect.~3.2 --- is found for a separation distance of 0.0574~AU,
and instability is found for a separation distance of 0.0576~AU.
The escape occurs after only 4442 years.  Furthermore, as indicated
in Figs. 6 and 7, short-term stability is found for retrograde
orbits for a separation distance of 0.1112~AU, whereas instability
is found for a separation distance of 0.1113~AU.  Here the escape
occurs after 3145 years.  Figure 8 shows an example of a long-term
($\simeq 10^6$ years) instability transition for an exomoon in a
retrograde orbit.  Here we chose two nearby starting positions,
i.e., 0.1107~AU and 0.1111~AU, to determine if the choice of
initial conditions in this region will largely determine the
long-term stability.

Figure 8 ({\it top panel}) depicts the orbit, within a rotating
reference frame, of an exomoon in a retrograde orbit using
the 0.1111~AU starting position; the exomoon escapes from the
system after 4000 years.  Furthermore, we investigated
how the semimajor axis ({\it middle panel}) and eccentricity
({\it bottom panel}) evolved over time considering these
two nearby initial positions.  Figure 8 ({\it middle panel})
shows that the semimajor axis for $a_m = 0.1107$~AU (depicted
as dots) remains relatively constant with minor fluctuations
in time for at least $10^5$ years, whereas the alternative $a_m$
(depicted as plus symbols) experiences an increase.  The
subsequent decrease in semimajor axis indicates the loss
of the moon due to ejection as also seen in the evolution
of eccentricity.

Figure 8 ({\it bottom panel}) demonstrates more clearly for
a $10^4$ year timespan the instability transition occurring
at 4000 years for the $a_m = 0.1111$~AU case, exhibiting
a similar trend in the value of eccentricity.  These simulations
were performed with the thoroughly tested GBS integration scheme.
Incidentally, in our previous study aimed at habitable Trojan
planets \citep{ebe11}, we already encountered a by-chance scenario
with a habitable moon in a retrograde orbit about the giant planet;
it occurred due to a particular set-up of the initial conditions.
In this case, the semimajor axis of the exomoon was identified
as 0.051~AU, which is consistent with the stability limit attained
in our present, more detailed study.

Technically speaking, the results on limits of
orbital stability described so far should be referred to
as {\it outer orbital stability limits}.  The {\it inner orbital
stability limits} of exomoons in the HD~23079 system are given by
the Roche limit of the giant planet; see discussion by \cite{wil03}
and references therein.  The evaluation of the Roche limit typically
yields a distance of about 2.45 radii from the central object
as stability limit; its assessment requires an approach
distinctly separate from the main focus of this paper.
The Roche limit is known to depend on, e.g., the radius,
oblateness, and density of the primary, and structural and
rotational properties of the satellite.  In the absence of
a more detailed discussion, the inner stability limit for
exomoons around HD~23079b can be estimated as about
0.0015~AU; note that this estimate is
weakly dependent on the mass of the giant planet.


\subsection{Statistical Assessments}

Another aspect of our investigation concerns the
statistical assessment of the orbital stability limits
obtained through numerical simulations.  Previously,
this has been done by applying the criterion of Hill
stability \citep[e.g.,][]{ham92,don10}, which often serves
as a customary approximation utilized in conjunction
with time-dependent simulations to gauge the onset of
orbital instability of planets in multi-body systems
\citep[e.g.,][]{men03,jon05,blo07}, although the
Hill stability concept has previously been identified
as largely defunct in particular applications \citep{cun09}.

For the parameter space as studied (see Tables 3 and 4)
we find that the stability limit for the exomoons varies
between 0.05236 and 0.06955~AU (prograde orbits) and between
0.1023 and 0.1190~AU (retrograde orbits) depending on the
orbital parameters of the Jupiter-type planet if a minimum mass
is assumed.  In general, we find that the larger the semimajor
axis $a_p$, the larger the stability limit.  Moreover, we also
find that the larger the eccentricity $e_p$, the smaller the
stability limit.  

Guided by the definition of the Hill radius (see Sect. 3.5),
we assume that the exomoon's orbital stability limit $z_{\rm lim}$
follows a relationship akin to
\begin{equation}
z_{\rm lim} \ = \ A_0 a_p^\alpha \Bigl(\frac{m_p}{3M}\Bigr)^\gamma (1-e_p)^\beta \ ,
\end{equation}
with $a_p$, $e_p$ and $m_p$ as defined.  The parameters
$A_0$, $\alpha$, $\beta$, and $\gamma$ are to be determined from the
data obtained by our orbital stability simulations.

First we focus on the case where the EGP is assumed to have minimum
mass, $m_p$; this allows us to constrain all parameters but $\gamma$
(see Table~6).  In this case, there are 25 models of prograde orbits
and 25 models of retrograde orbits available.  Fitting the data
yields $A_0 = 0.296$, $\alpha = 1.02$, and $\beta = 0.68$ for
the prograde models, and $A_0 = 0.572$, $\alpha = 0.99$, and
$\beta = 0.52$ for the retrograde models.  As quality
checks we assessed the mean and maximum errors obtained for the
two data sets as well as the errors for the master models (MMod)
given as $a_p = 1.596$~AU and $e_p = 0.102$.

We find that these fits for the prograde and retrograde cases are
fully acceptable; the errors are typically $\lta 10\%$ while
noting that in the retrograde case the errors are considerably
smaller as some stability limits in the prograde case are adversely
affected by resonance phenomena (see below).  In order to simplify
the fits, we pursued tests with $\alpha = 1.0$ and $\beta = 0.5$
as well as $\beta = 1.0$ for the prograde and retrograde cases.
Even though these choices were found to be largely inconsequential
for the quality of the fits, we elected $\alpha = 1.0$ and
$\beta = 1.0$ (adjusted fit~2) as our preferred choice for both
the prograde and retrograde cases, as it also reflects the
functional form of the Hill stability radius (see Sect. 3.5).
Additionally, it is found that the preferred fit renders the
lowest maximum and MMod errors compared to the best fit and the
adjusted fit~1.  Moreover, the differences regarding the mean errors
between adjusted fit~2 (our preferred choice) and the best fit
are also found to be relatively minor.

Some values in Table 3 do not perfectly fit the expected progression of 
stability limit regarding the surrounding ($a_p$, $e_p$) results.  
Specifically, we find that for the cases where $a_p$ = 1.6425, the predicted 
stability limits would reside near an 11:2 mean motion resonance.  Recent 
studies with the circumbinary planet, Kepler-16b, has shown that the Saturnian 
planet routinely clears the surrounding stability pockets near the same 
resonance \citep{qua12,pop12}.  Thus our estimations of the stability limit
for this region would be truncated due to this type of behavior.  Fortunately,
our overall statistics is not considerably impacted by this phenomenon.

As part of our study, the mass of the EGP was increased by factors of
1.5 and 2.0 relative to its minimum mass (see Table~5).  This also allows
us to deduce the parameter $\gamma$ (see Eq.~1), which we identified as
$\gamma = 0.309 \pm 0.061$ (1$\sigma$) with $\sigma_{\rm M} = 0.031$,
the standard error of the mean.  Note that the value of $\gamma$
as determined does not significantly deviate from the theoretical value
of $\gamma = 1/3$; it obviously agrees with the model-dependent value
by less than 1$\sigma_{\rm M}$.

Another quantity of interest concerning our stability studies
is the ratio of orbital stability limits between models
of moons in retrograde and prograde orbits, i.e.,
\begin{equation}
\eta \ =  \ z_{\rm lim}^{\rm retrograde} / z_{\rm lim}^{\rm prograde} .
\end{equation}
We find that $\eta$ is given as $1.891 \pm 0.094$ for models with EGP's
minimum mass; here $\sigma_{\rm M}$ is given as 0.019.  If the four
models with the increased mass of the EGP are also included, $\eta$
is identified as $1.902 \pm 0.098$ with $\sigma_{\rm M} = 0.019$.
Thus, in conclusion, the stability limit for exomoons in
retrograde orbits is increased by 90\% over the stability limit
for prograde orbits.


\subsection{Comparison to Previous Work}

Finally, we can also compare the stability limits obtained through
our numerical simulations to the results of the fitting
formulae given by \cite{dom06} (D06).  With $R_{\rm H}$ denoting
the well-established Hill radius \citep[e.g.,][]{ham92}, given as
\begin{equation}
R_{\rm H} \ \simeq \ a_p \Bigl(\frac{m_p}{3M}\Bigr)^{1/3} (1-e_p) \ ,
\end{equation}
where $a_p$ and $e_p$ are the semi-major axis and eccentricity of the
planetary orbit, respectively, and $M$ and $m_p$ are the masses of the
star and planet, respectively, the limits of stability for prograde
and retrograde orbits of the Earth-mass exomoon are given as
\begin{equation}
z_{\rm lim}^{\rm prograde} \ = \ 0.4895 \ (1.0 - 1.0305 e_p - 0.2738 e_m) R_{\rm H}
\end{equation}
and
\begin{eqnarray}
z_{\rm lim}^{\rm retrograde}  &= \ 0.9309 \ (1.0 - 1.0764 e_p - 0.9812 e_m \nonumber \\
                              &    + 0.9446 e_p e_m) R_{\rm H} \ , \phantom{XXXX}
\end{eqnarray}
respectively, with $e_m$ denoting the orbital eccentricity of the exomoon.

A simple test of using the D06 formulae for the limiting case of $e_p \rightarrow 0$
and $e_m \rightarrow 0$ reveals a promising result: the value of $\eta$ (D06) is
identified as $0.9309 / 0.4895 = 1.902$, which is in perfect agreement
with the $\eta$ value identified for our simulations (see Sect. 3.4).

However, a more appropriate evaluation of the D06 formulation requires the
checking of its applicability to the $z_{\rm lim}$ values for prograde and
retrograde orbits in a separate manner.  Thus, we introduce
\begin{equation}
q \ = \ z_{\rm lim}^{\rm D06} / z_{\rm lim}^{\rm this~work} \ .
\end{equation}
In this regard we find a significantly different outcome for prograde
and retrograde models.  If $e_m = 0$ is assumed, no agreements between
our numerical simulations and the results of the D06 formulae are attained.
However, during orbital instability transitions, the orbital
eccentricity of the exomoon is expected to be relatively large, i.e., 
$e_m \rightarrow 1$; see, e.g., the analysis by \cite{ebe10a}.

Therefore, for prograde orbits, if $e_m = 0.95$ is used, we find $q = 1.027$
with $\sigma_{\rm M} = 9.11 \times 10^{-3}$.  Moreover, if $e_m = 0.99$ is
used, we find $q = 1.005$ with $\sigma_{\rm M} = 9.06 \times 10^{-3}$.
Hence, for prograde orbits, the D06 formula is
identified as fully applicable to represent the $z_{\rm lim}$ values of
orbital stability for exomoons in the HD~23079 system.  However, similar
tests for retrograde orbits show that the D06 formula is off by about
an order of magnitude.  Thus, we conclude that the D06 formulation has not
been well-tested for parameter sets of the latter kind.


\section{Summary and Conclusions}

The aim of our study was to explore the general possibility of habitable
moons in the HD~23079 star-planet system.  This system consists of
a main-sequence star slightly hotter than the Sun.  It also
contains a Jupiter-mass planet, HD~23079b, with a minimum mass of
2.45~$M_J$ in a slightly elliptical orbit that is almost certainly
completely positioned in HD~23079's HZ.  We studied if
Earth-mass habitable moons can potentially exist in this system.  Thus,
we deduced the orbital stability limits of moons orbiting HD~23079b by
considering both prograde and retrograde orbits.

The set of models of our study also took into account the
observational uncertainties of $a_p$ and $e_p$ for HD~23079b, which
for the minimum mass case of the EGP resulted in the establishment
of a $5 \times 5$ grid of stability limits for both prograde and
retrograde orbits; clearly, a much larger number of models was
pursued for the derivation of the stability limits due to the
required test simulations given by different exomoon
starting distances from the EGP.  Moreover, the semimajor axis
$a_p$ of the EGP was varied from 1.503 to 1.689~AU in 0.0465 increments,
whereas its eccentricity $e_p$ was varied from 0.071 to 0.133 in 0.0155
increments.  Furthermore, for the master model with $a_p = 1.596$~AU
and $e_p = 0.102$, the mass of the EGP, $m_p$, was increased by
factors of 1.5 and 2.0.

For the minimum mass case of the EGP, the orbital stability limits
were found to vary between 0.05236 and 0.06955~AU (prograde orbits)
and 0.1023 and 0.1190~AU (retrograde orbits), corresponding
0.306 and 0.345 (prograde orbits) and 0.583 and 0.611 (retrograde
orbits) of the EGP's Hill radius.  Somewhat larger stability
limits for the exomoons were obtained if an increased value for the
mass of the Jupiter-type planet was adopted, as expected.  Generally,
it was found that the larger the semimajor axis $a_p$, the larger
the stability limit.  It was also found that the larger the eccentricity
$e_p$, the smaller the stability limit.  A small number of exceptions
was identified when the orbital motion of the exomoon was affected by
resonances.  Also note that the limits as deduced constitute, technically
speaking, outer orbital stability limits; the inner limits of orbital
stability are, in essence, given by the EGP Roche limit.

Another element of our study was the derivation of the functional
dependencies of the orbital stability limits on the EGP parameters
based on $z_{\rm lim} \propto a_p^\alpha m_p^\gamma (1-e_p)^\beta$
guided by the customary notion of Hill stability.  Based on our
simulations, we adopted $\alpha = 1$, $\beta = 1$, and $\gamma = 1/3$
for both prograde and retrograde orbits, which is in agreement with
the functional dependencies of the Hill stability radius.  In addition,
we also found that the stability limits for prograde orbits
were in perfect agreement with the analytical formula of
\cite{dom06}.  However, no such agreement was identified for
retrograde orbits, which is most likely attributable to that the
formula for retrograde satellite motion of \cite{dom06} may be
inapplicable for $e_p \rightarrow 0$ and $e_m \rightarrow 1$.

A further quantity considered in our study concerns
the ratio of orbital stability limits between models of moons
in retrograde and prograde orbits $\eta$ (see Eq.~2).
According to our model simulations, we found that $\eta$ is
given as 1.902 with $\sigma_{\rm M} = 0.019$.  Thus, on average,
the stability limit for retrograde orbits is increased by
90\% compared to the limit of stability for prograde orbits.
This result is, in principle, consistent with the previous
study by \cite{jef74} who explored the orbital stability of
the restricted 3-body problem based on the usage of Henon
stability maps.  He identified considerably broader stability
regions for the massless particle-type object if a retrograde
instead of a prograde orbit about the primary object was assumed.

Another example for significantly increased stability limits
for retrograde orbits has been given by \cite{ebe10b}.  They
pursued a theoretical investigation of a previous observational
finding about the $\nu$~Octantis binary system \citep{ram09} that
indicates the possible existence of a Jupiter-mass planet, although
the planet appears to be located outside the zone of orbital
stability.  Based on detailed numerical studies, \cite{ebe10b}
argued that the Jupiter-mass may be in a retrograde orbit
relative to the motion of the binary components, which would
allow the planet to be orbitally stable.  Additional results
in line with this reasoning have recently been given by
\cite{qua12} and \cite{goz13}.

The study of exomoons is currently
receiving heightened attention because of the
ongoing Kepler mission.  For example, recent work by \cite{kip09}
addressed the detectability of moons of transiting planets,
including those that are expected to be detected by Kepler or
photometry of approximately equal quality.  These studies
convey both the predicted transit timing signal amplitudes
and the estimated uncertainties on such measurements to provide
realistic confidence levels for the detection of such bodies
for a broad spectrum of orbital arrangements.  They
concluded that HZ exomoons down to $0.2~M_\oplus$ may be
detectable and approximately 25000 stars could be surveyed
for HZ exomoons with Kepler's field of view.  Moreover,
\cite{kip13} recently presented seven possible candidates for
transiting exomoons using observations from the Kepler mission.
Previously, \cite{kal10} discussed the possibility of screening
the atmospheres of exomoons for habitability based on the
concept of biomarkers.  These investigations will potentially
allow identifying the existence of habitable exomoons around
different types of stars, which will allow contesting
results and predictions from earlier studies by \cite{wil97},
\cite{bar02}, and more recently by \cite{kip09}, \cite{kal10},
\cite{wei10}, \cite{rob12}, \cite{hel13}, and others.

This study complements a previous investigation by our research
group, given by \cite{ebe11}, that showed that the HD~23079
star--planet system is also a suitable candidate for hosting
habitable Trojan planets.

\begin{figure}
\centering
\includegraphics[height=7cm]{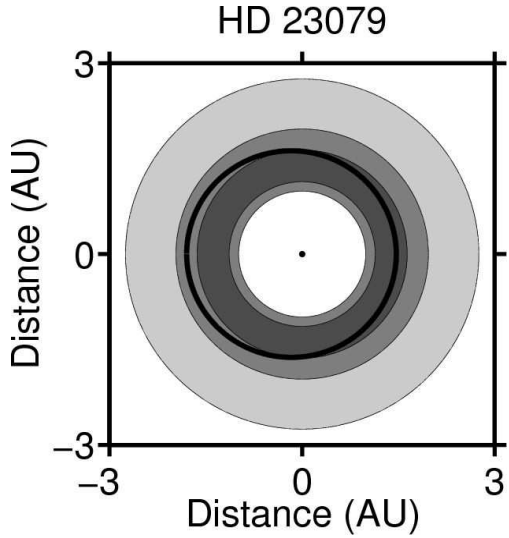}
\caption{
Extent of the HZ for HD~23079, defined by its conservative limits
(dark gray) and generalized limits (medium gray).  In addition, we
depict the outer limit of an extreme version of the generalized HZ
(light gray) following the work by \cite{mis00}, although this limit
may still be controversial.  The orbit of HD~23079b is depicted by
a thick solid line.
}
\bigskip
\end{figure}

\begin{figure*}
\centering
\includegraphics[height=6cm]{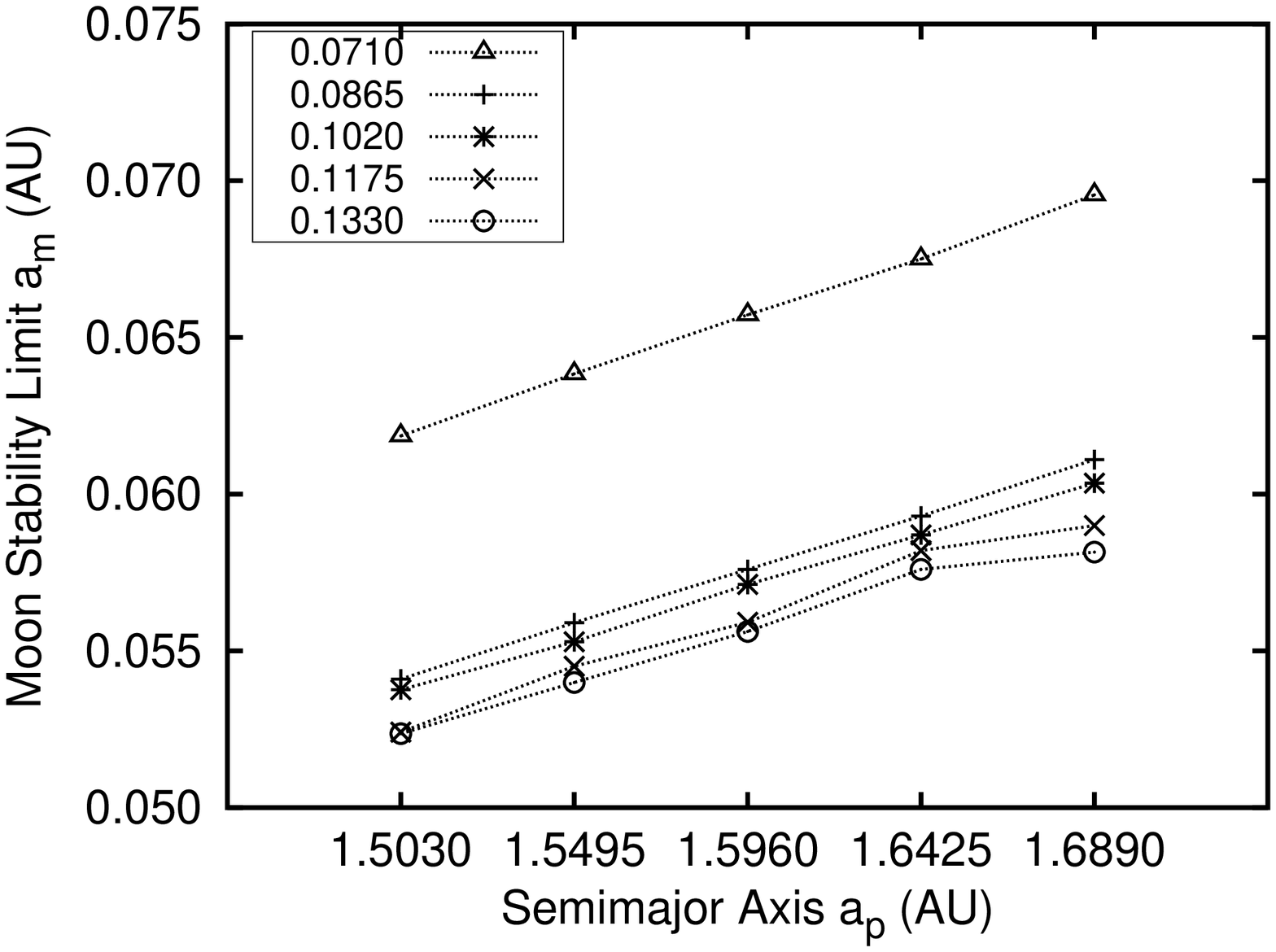}
\caption{
Depiction of orbital stability limits of habitable moons for prograde
orbits; see Table~3 for data information.  The data are grouped in regard
to the EGP's eccentricity $e_p$.
}
\bigskip
\end{figure*}

\begin{figure*}
\centering
\includegraphics[height=6cm]{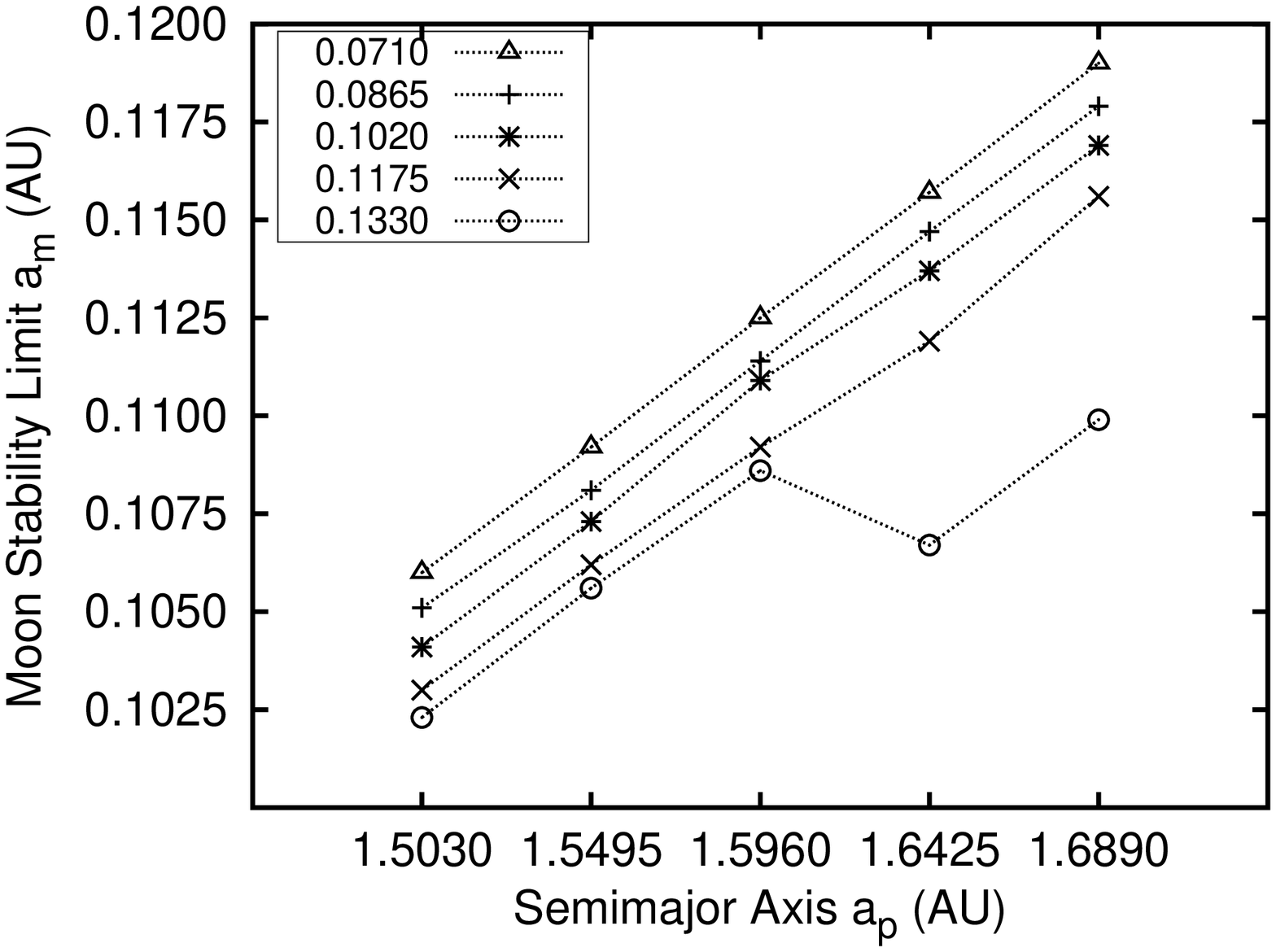}
\caption{
Depiction of orbital stability limits of habitable moons for retrograde
orbits; see Table~4 for data information.  The data are grouped in regard
to the EGP's eccentricity $e_p$.
}
\bigskip
\end{figure*}

\begin{figure*}
\centering
\includegraphics[height=6cm]{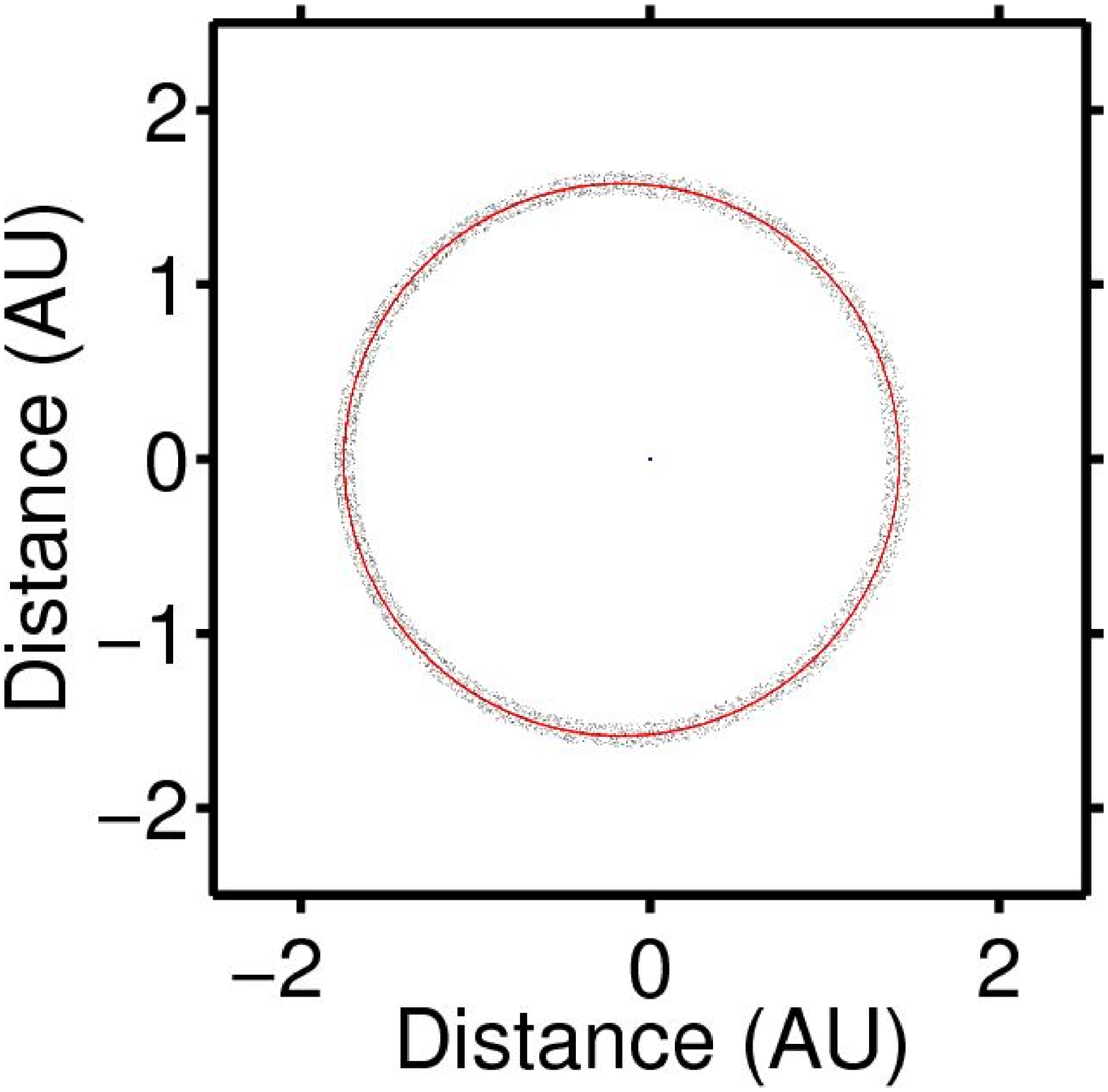}
\includegraphics[height=6cm]{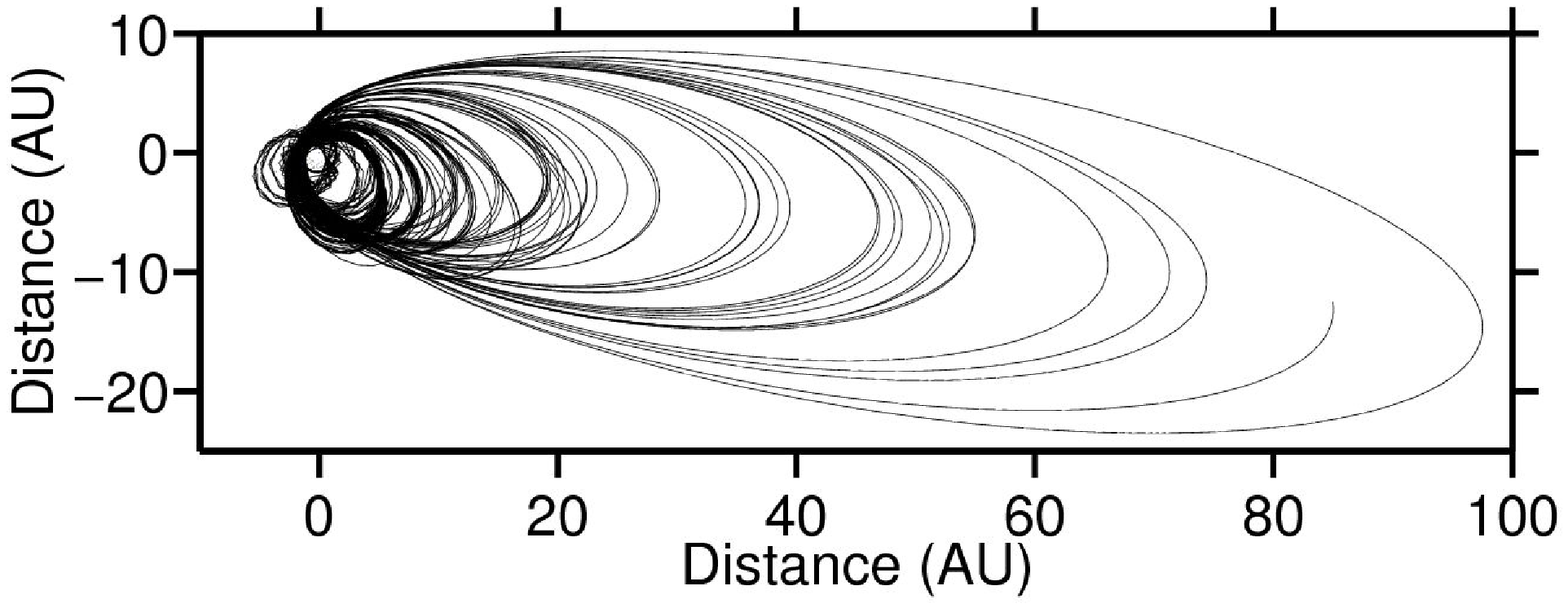}
\caption{
Orbital stability simulations for the moon assumed to orbit
HD~23079b (red line) for an elapsed simulation time of
10$^4$ years.  The orbital
parameters of the planet are given as $a_p = 1.596$~AU and
$e_p = 0.102$.  The moon has been placed in a prograde orbit
about the planet with its starting distance given as 0.0574~AU
(top) and 0.0576~AU (bottom).
}
\bigskip
\end{figure*}

\begin{figure}
\centering
\includegraphics[height=6cm]{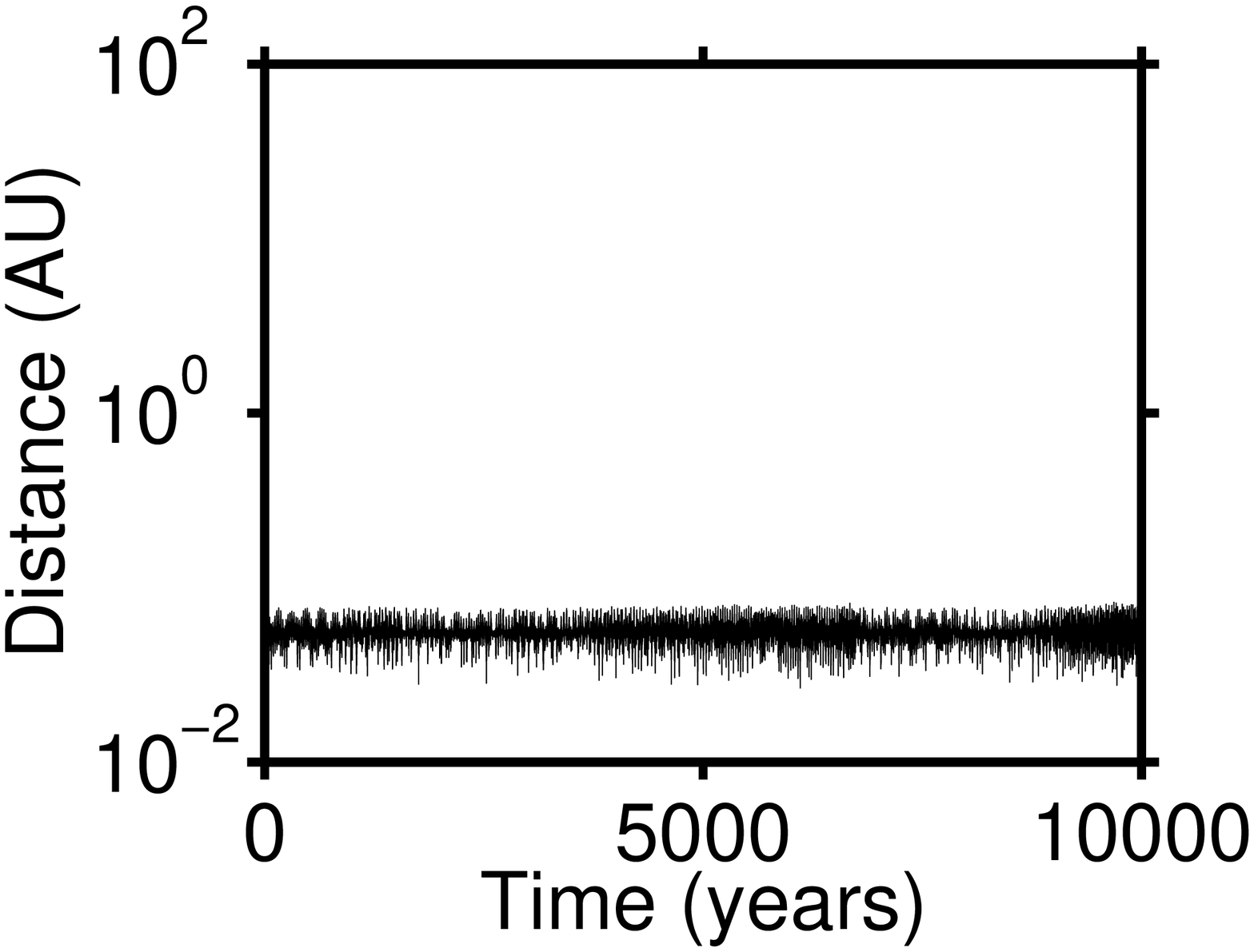} \\
\includegraphics[height=6cm]{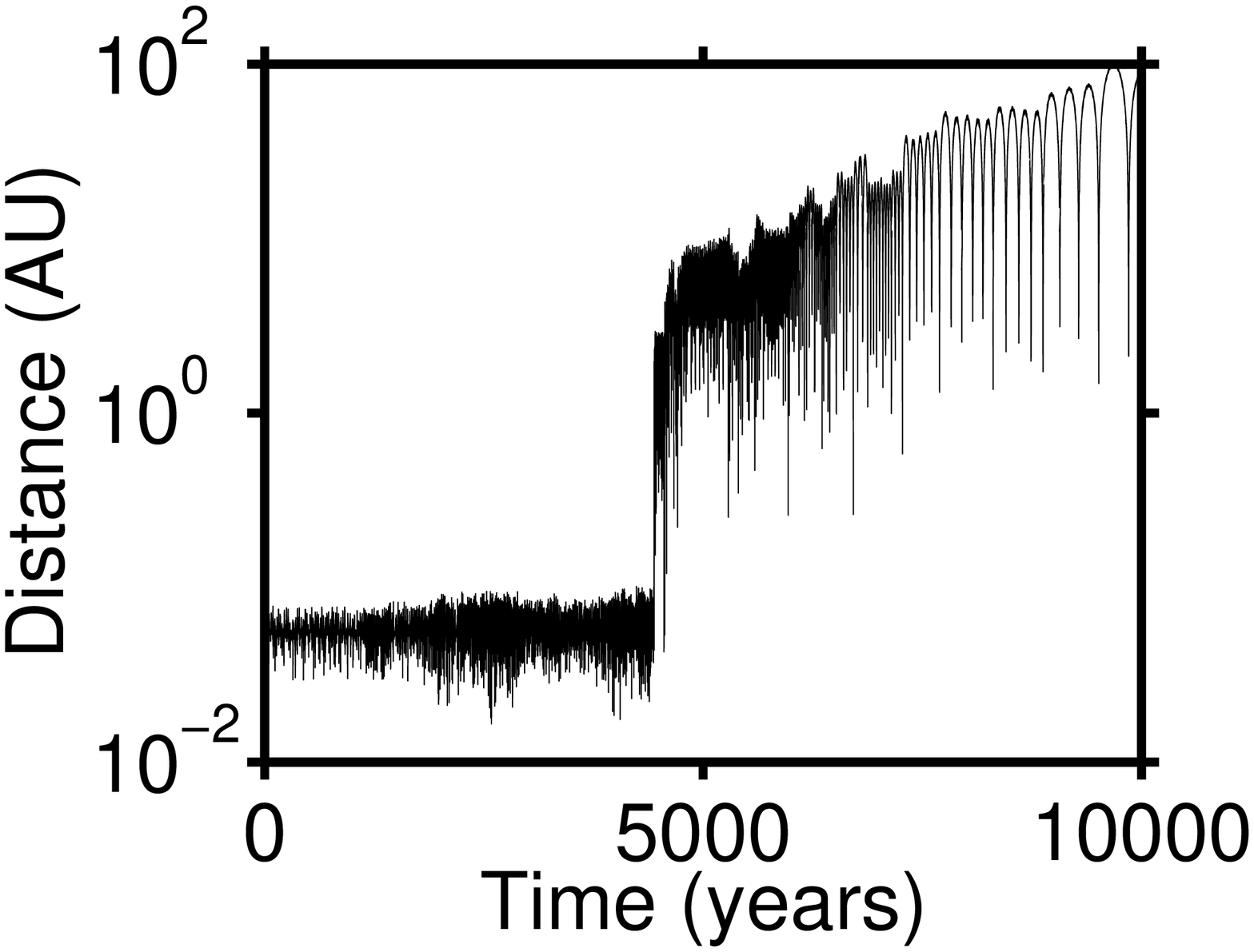}
\caption{
Separation distance of the moon from HD~23079b for the orbital stability
study depicted in Fig.~4.  Note that the moon has been placed in a
prograde orbit about the planet with its starting distance given as
0.0574~AU (top) and 0.0576~AU (bottom).
}
\bigskip
\end{figure}

\begin{figure}
\centering
\includegraphics[height=6cm]{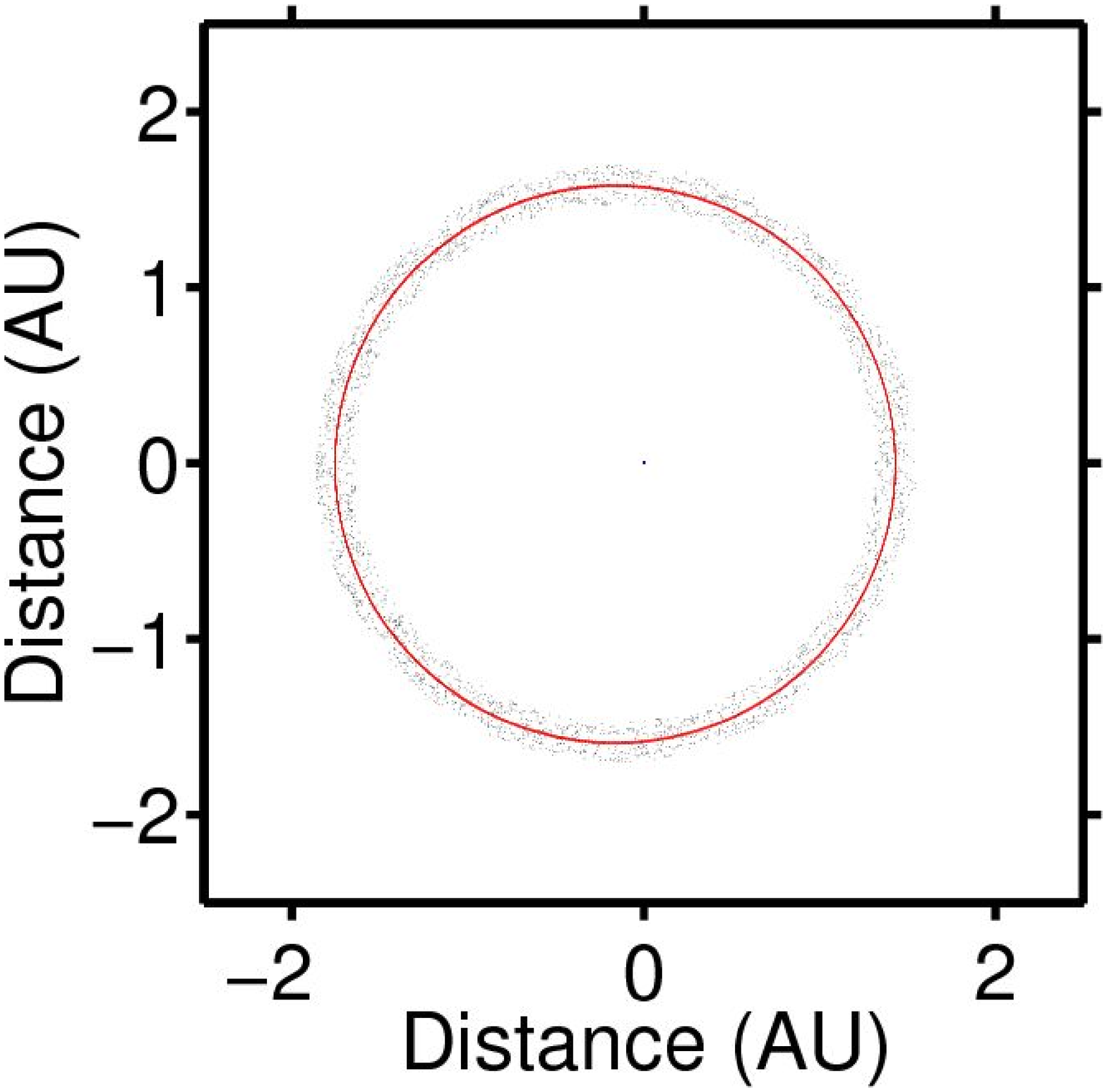} \\
\includegraphics[height=6cm]{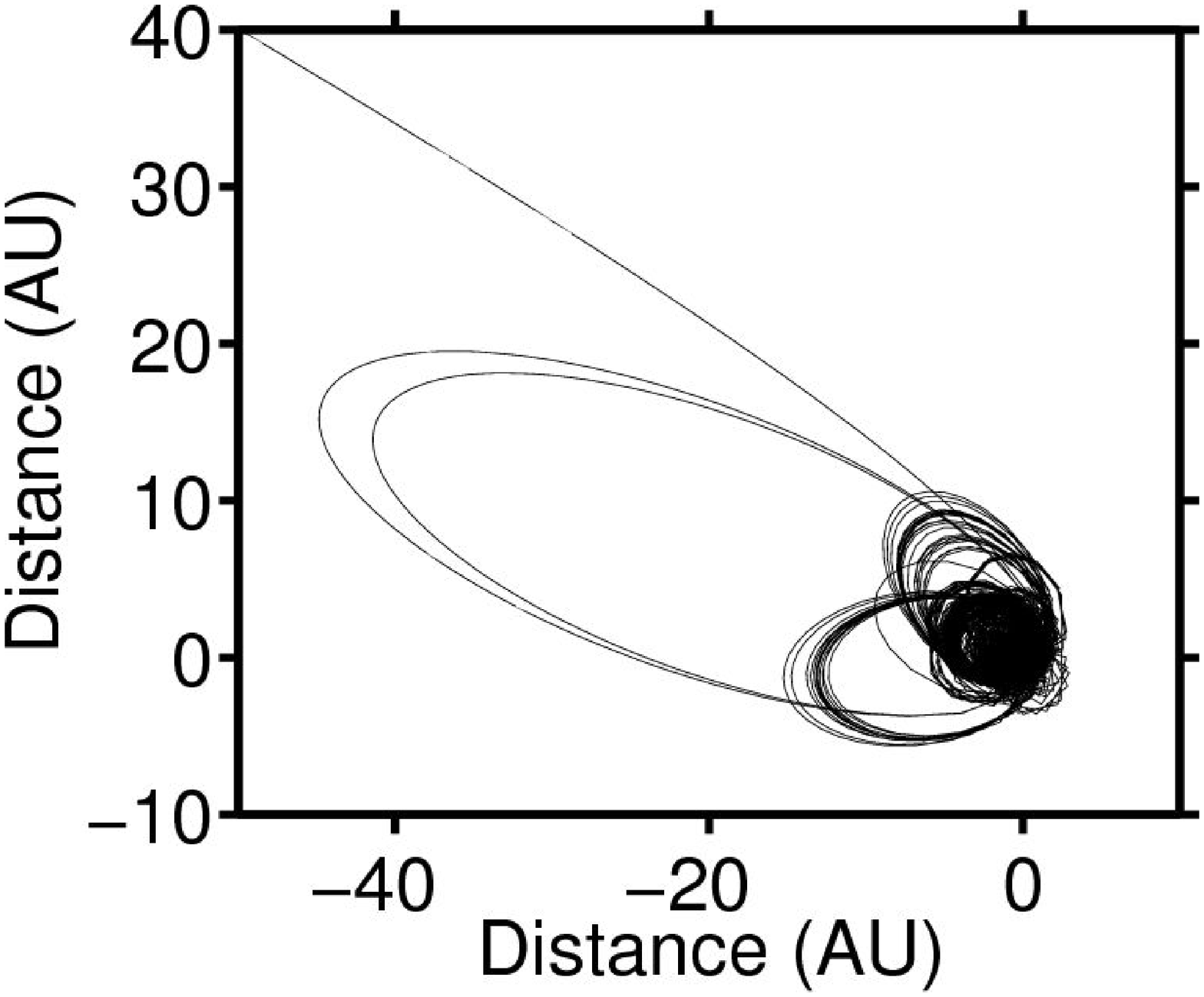}
\caption{
Orbital stability simulations for the moon assumed to orbit
HD~23079b (red line) for an elapsed simulation time of
10$^4$ years.  The orbital
parameters of the planet are given as $a_p = 1.596$~AU and
$e_p = 0.102$.  The moon has been placed in a retrograde orbit
about the planet with its starting distance given as 0.1112~AU
(top) and 0.1113~AU (bottom).
}
\bigskip
\end{figure}

\begin{figure}
\centering
\includegraphics[height=6cm]{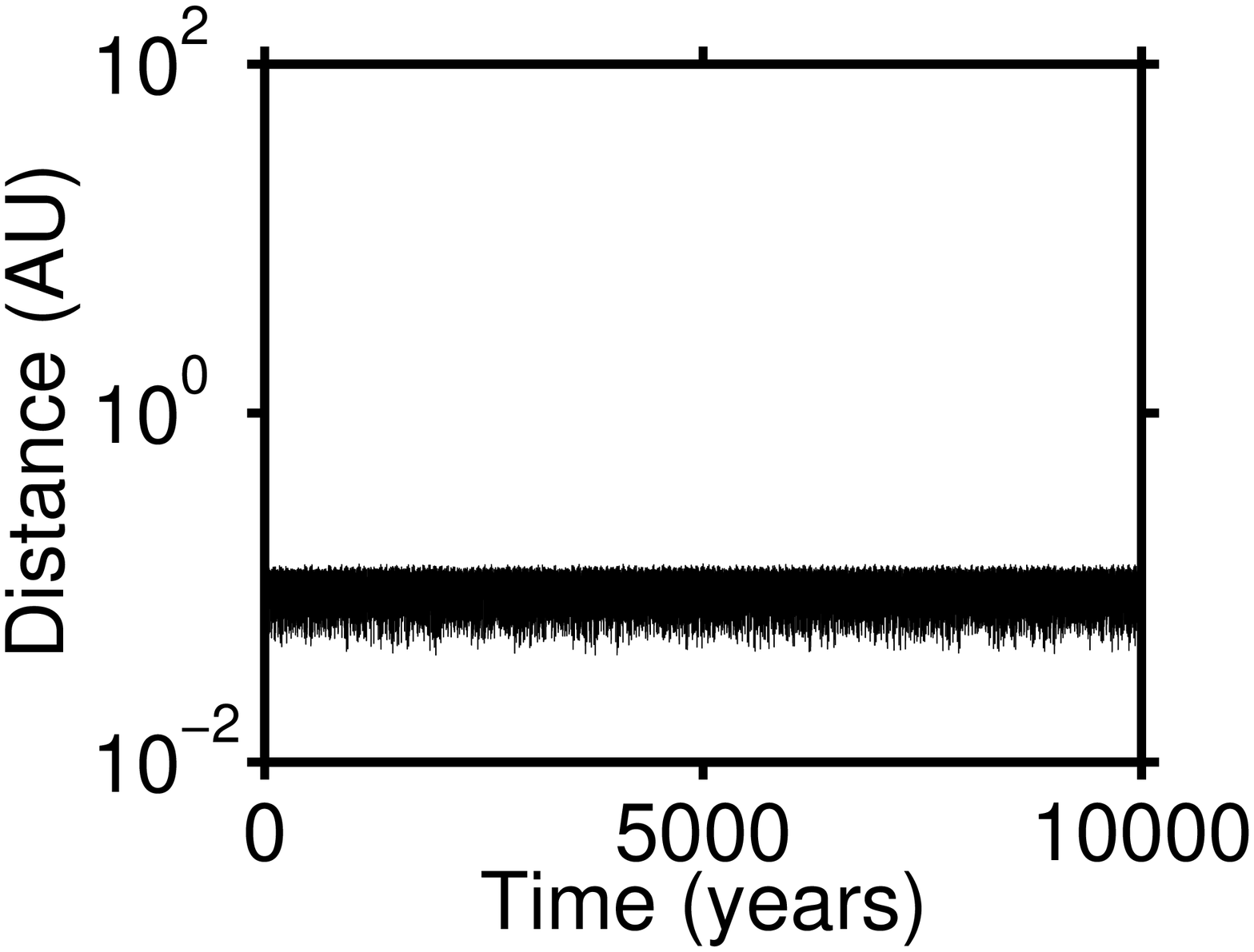} \\
\includegraphics[height=6cm]{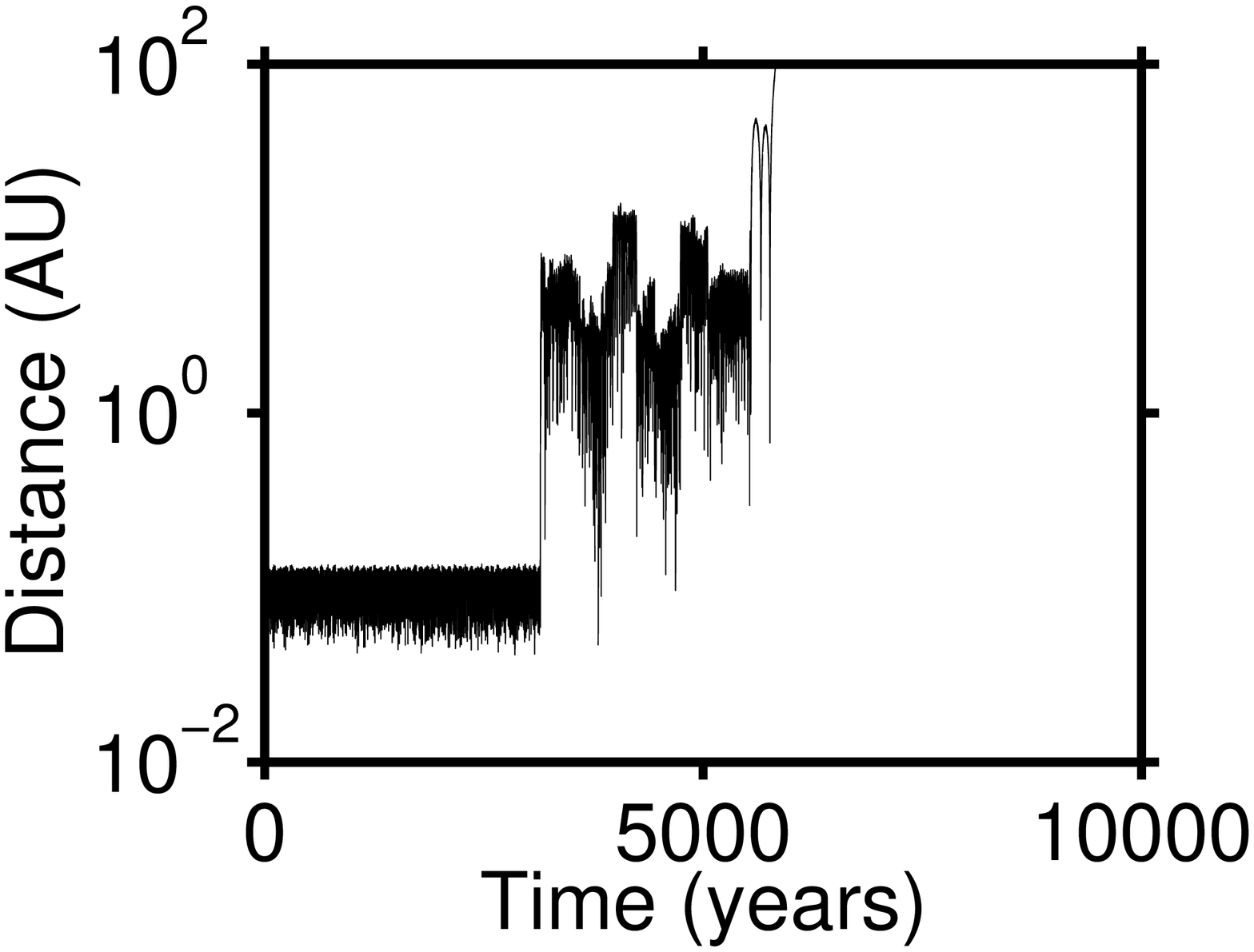}
\caption{
Separation distance of the moon from HD~23079b for the orbital stability
studies depicted in Fig.~6.  Note that the moon has been placed in a
retrograde orbit about the planet with its starting distance given as
0.1112~AU (top) and 0.1113~AU (bottom).
}
\bigskip
\end{figure}

\begin{figure}
\centering
\includegraphics[scale=0.15]{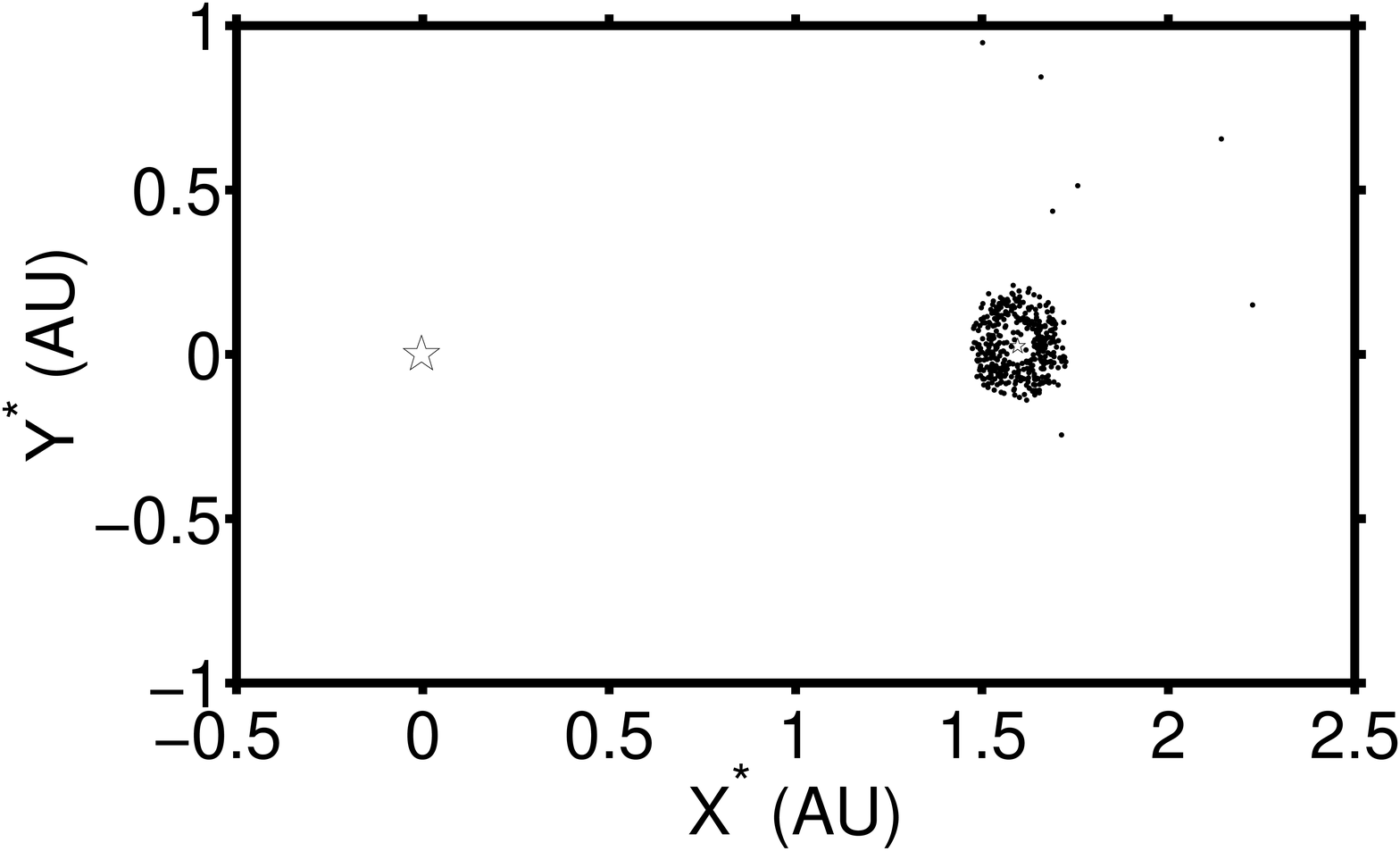} \\
\includegraphics[scale=0.15]{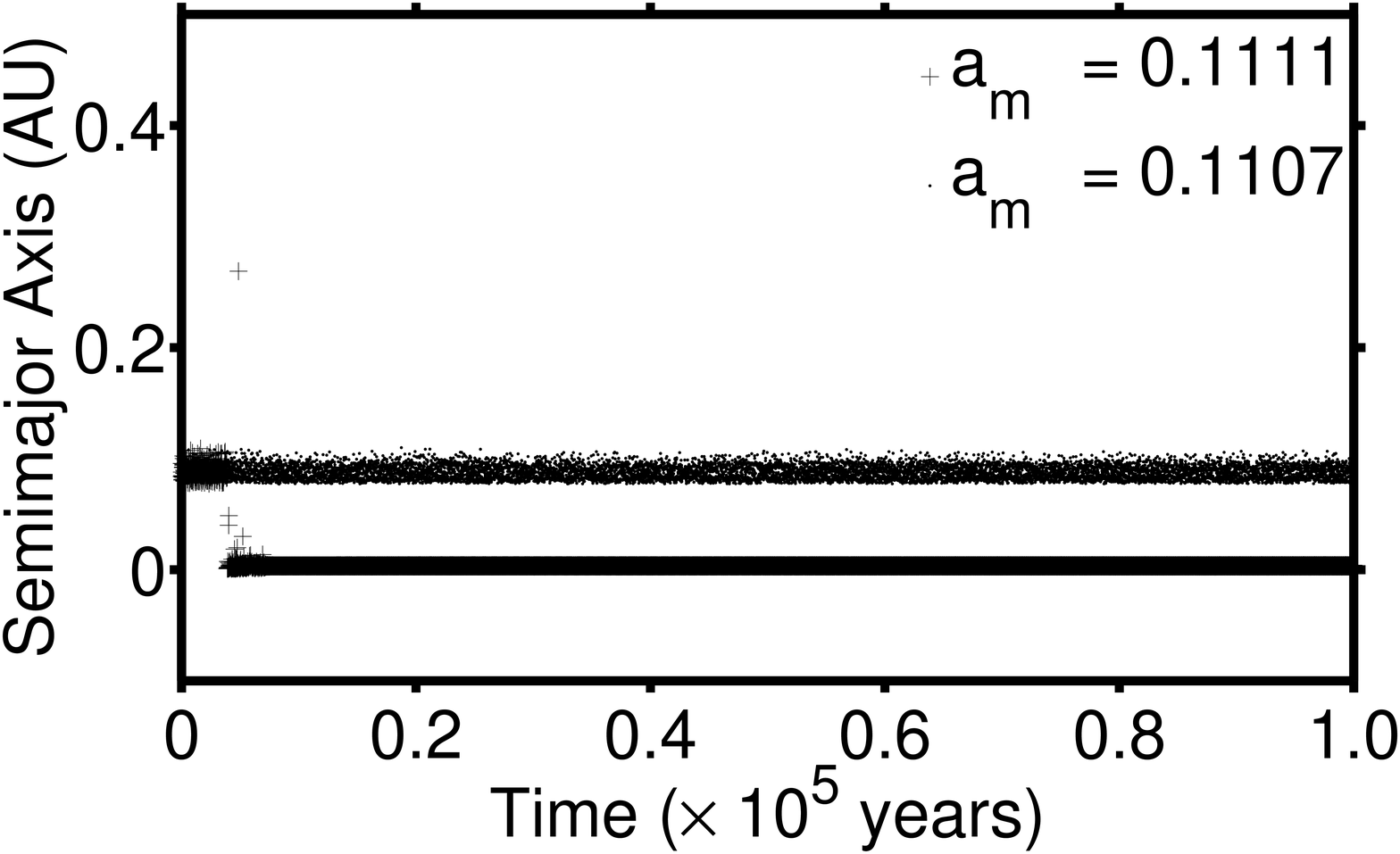} \\
\includegraphics[scale=0.15]{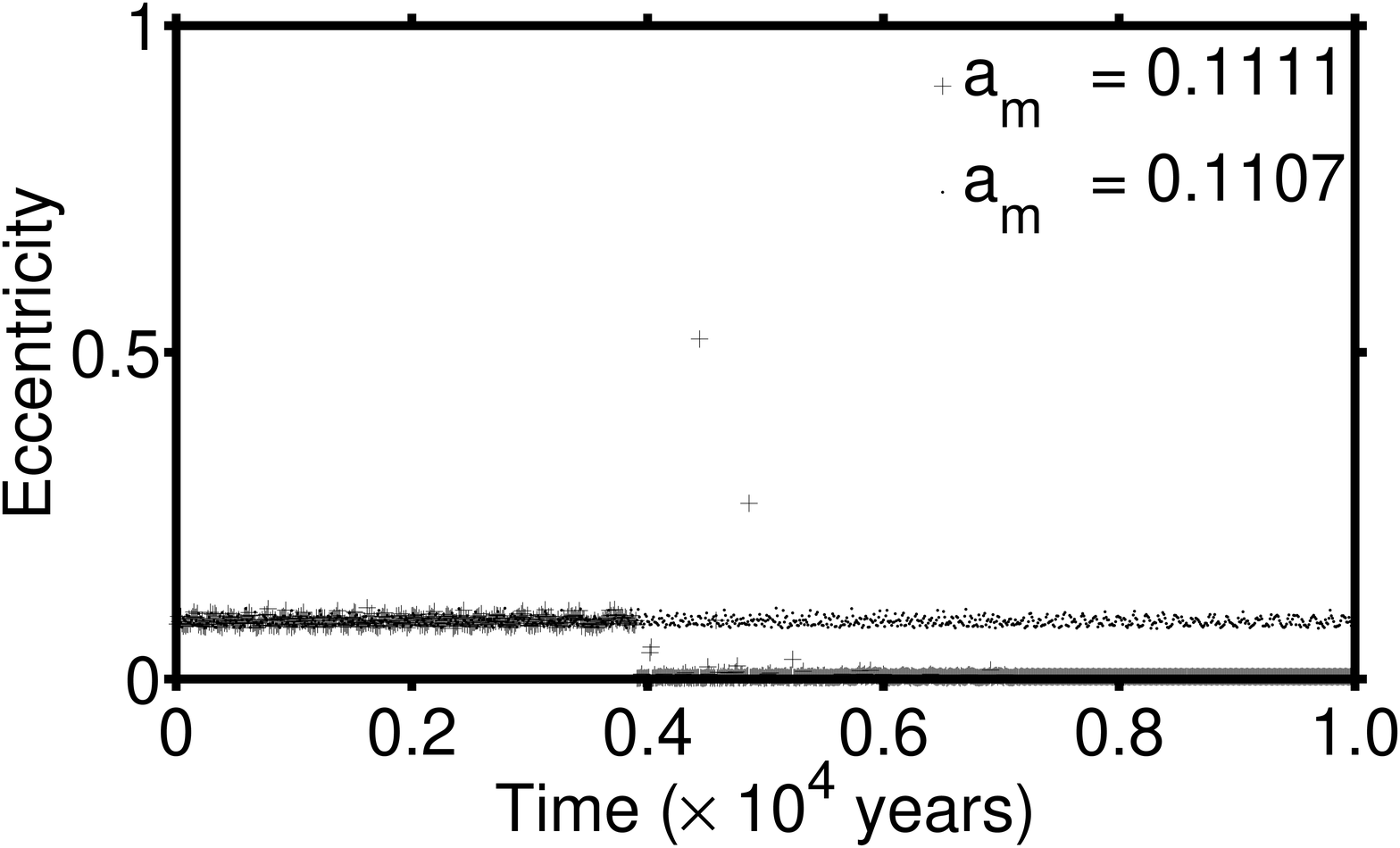}
\caption{
Orbital stability simulations for the moon assumed to orbit
HD~23079b (small five-side star).  The orbital parameters of the EGP
are given as $a_p = 1.596$~AU and $e_p = 0.102$.  The moon has been
placed in a retrograde orbit within a rotating reference frame
about the planet with its starting distance given as 0.1111~AU
(top). Time series of the semimajor axis (middle) and eccentricity
(bottom) are shown with respect to the designated starting distance
in a retrograde orbit.
}
\bigskip
\end{figure}

\begin{table*}
\bigskip
\caption{Stellar and Planetary Parameters}
\bigskip
\centering
\begin{tabular}{l c c}
\noalign{\smallskip}
\hline
\hline
\noalign{\smallskip}
Parameter               &  Value                                              & Reference \\
\noalign{\smallskip}
\hline
\noalign{\smallskip}
 Spectral Type          &    F9.5~V                                           & \cite{gra06}   \\
 RA                     &       3$^{\rm h}$ 39$^{\rm m}$ 43.0952$^{\rm s}$    & \cite{esa97}$^{a,b}$ \\
 DEC                    &  $-$ 52$^\circ$ 54$^\prime$ 57.017$^{\prime\prime}$ & \cite{esa97}$^{a,b}$ \\
 $T_{\rm eff}$~(K)      &  6030  $\pm$ 52                                     & \cite{rib03}   \\
 $R$~($R_\odot$)        &  1.106 $\pm$ 0.022                                  & \cite{rib03}   \\
 $M$~($M_\odot$)        &  1.10  $\pm$ 0.15                                   &    $^c$        \\
 $M_V$                  &  4.42  $\pm$ 0.05                                   & \cite{esa97}$^{a,b,d}$ \\
 $M_{\rm bol}$          &  4.25  $\pm$ 0.05                                   & \cite{esa97}$^{a,b,d}$ \\
 Distance~(pc)          & 34.60  $\pm$ 0.67                                   & \cite{esa97}$^{a,b,d}$ \\
 $m_p {\sin}i$~($M_J$)  &  2.45  $\pm$ 0.21                                   & \cite{but06}   \\
 $P$~(days)             & 730.6  $\pm$ 5.7                                    & \cite{but06}   \\
 $a_p$~(AU)             &  1.596 $\pm$ 0.093                                  & \cite{but06}   \\
 $e_p$                  &  0.102 $\pm$ 0.031                                  & \cite{but06}   \\
\noalign{\smallskip}
\hline
\end{tabular}
\medskip \\
$^a$data from SIMBAD, see {\tt http://simbad.u-strasbg.fr} \\
$^b$adopted from the {\it Hipparcos} catalogue \\
$^c$based on spectral type \\
$^d$based on parallax $28.90 \pm 0.56$~mas
\end{table*}
\noindent

\begin{table*}
\bigskip
\caption{Orbit of the EGP and HZ}
\bigskip
\centering
\begin{tabular}{l c c}
\noalign{\smallskip}
\hline
\hline
\noalign{\smallskip}
Parameter       &  Value   & $\sigma$ \\
\noalign{\smallskip}
\hline
\noalign{\smallskip}
 $a_{\rm per}$  &  1.432   & 0.094   \\
 $a_{\rm ap}$   &  1.758   & 0.113   \\
 HZ-i (gen.)    &  0.989   & 0.027   \\
 HZ-i (cons.)   &  1.138   & 0.031   \\
 HZ-o (cons.)   &  1.636   & 0.044   \\
 HZ-o (gen.)    &  1.966   & 0.054   \\
\noalign{\smallskip}
\hline
\end{tabular}
\medskip \\
Note: All data are given in units of AU.
\end{table*}

\begin{table*}
\bigskip
\caption{Stability Limits of Habitable Moons; Prograde Orbits}
\bigskip
\centering
\begin{tabular}{l c c c c c}
\noalign{\smallskip}
\hline
\hline
\noalign{\smallskip}
$e_p$     &  \multicolumn{5}{c}{$a_p$}   \\
\noalign{\smallskip}
\hline
\noalign{\smallskip}
...       &  1.5030  &  1.5495  &  1.5960  &  1.6425  &  1.6890 \\
...       &  (AU)    &  (AU)    &  (AU)    &  (AU)    &  (AU)   \\
\noalign{\smallskip}
\hline
\noalign{\smallskip}
   0.0710  &  0.06186  &  0.06384  &  0.06572  &  0.06750  &  0.06955 \\
   0.0865  &  0.05410  &  0.05590  &  0.05760  &  0.05930  &  0.06110 \\
   0.1020  &  0.05376  &  0.05529  &  0.05712  &  0.05870  &  0.06035 \\
   0.1175  &  0.05241  &  0.05451  &  0.05591  &  0.05820  &  0.05900 \\
   0.1330  &  0.05236  &  0.05399  &  0.05562  &  0.05760  &  0.05815 \\
\noalign{\smallskip}
\hline
\end{tabular}
\medskip \\
Note: $a_p$ is given in units of AU.  The stability limits represent
maximum radii for the orbits.
\end{table*}

\begin{table*}
\bigskip
\caption{Stability Limits of Habitable Moons; Retrograde Orbits}
\bigskip
\centering
\begin{tabular}{l c c c c c}
\noalign{\smallskip}
\hline
\hline
\noalign{\smallskip}
$e_p$     &  \multicolumn{5}{c}{$a_p$}   \\
\noalign{\smallskip}
\hline
\noalign{\smallskip}
...       &  1.5030  &  1.5495   &  1.5960  &  1.6425   &  1.6890 \\
...       &  (AU)    &  (AU)     &  (AU)    &  (AU)     &  (AU)   \\
\noalign{\smallskip}
\hline
\noalign{\smallskip}
   0.0710 &  0.1060  & 0.1092   & 0.1125   & 0.1157   & 0.1190 \\
   0.0865 &  0.1051  & 0.1081   & 0.1114   & 0.1147   & 0.1179 \\
   0.1020 &  0.1041  & 0.1073   & 0.1109   & 0.1137   & 0.1169 \\
   0.1175 &  0.1030  & 0.1062   & 0.1092   & 0.1119   & 0.1156 \\
   0.1330 &  0.1023  & 0.1056   & 0.1086   & 0.1067   & 0.1099 \\
\noalign{\smallskip}
\hline
\end{tabular}
\medskip \\
Note: $a_p$ is given in units of AU.  The stability limits represent
maximum radii for the orbits.
\end{table*}

\begin{table*}
\bigskip
\caption{Stability Limits of Habitable Moons for Different EGP Masses}
\bigskip
\centering
\begin{tabular}{l c c}
\noalign{\smallskip}
\hline
\hline
\noalign{\smallskip}
Mass of EGP  &  Prograde Orbit &  Retrograde Orbit \\
\noalign{\smallskip}
\hline
\noalign{\smallskip}
           $1.0 \times m_p$ & 0.05712  &   0.1109 \\
           $1.5 \times m_p$ & 0.06290  &   0.1291 \\
           $2.0 \times m_p$ & 0.06940  &   0.1400 \\
\noalign{\smallskip}
\hline
\end{tabular}
\medskip \\
Note: The simulations are given for $a_p = 1.596$~AU and $e_p = 0.102$.
All data are given in units of AU.  The stability limits represent
maximum radii for the orbits.
\end{table*}

\begin{table*}
\bigskip
\caption{Functional Parameters of $z_{\rm lim}$, Error Analysis}
\bigskip
\centering
\begin{tabular}{l l c c c c c c}
\noalign{\smallskip}
\hline
\hline
\noalign{\smallskip}
Model  & Type & A$_0$  & $\alpha$ & $\beta$ &  Mean Error &  Max. Error &  MMod Error \\
...    & ...  & ...    & ...      & ...     &  \%         &  \%         &  \%         \\
\noalign{\smallskip}
\hline
\noalign{\smallskip}
 Prograde Orbits   & best fit        &  0.296  &  1.02  &  0.68  &  2.5  &  11.4  &  3.2   \\
 Prograde Orbits   & adjusted fit 1  &  0.293  &  1.00  &  0.50  &  2.8  &  11.9  &  3.7   \\
 Prograde Orbits   & adjusted fit 2  &  0.309  &  1.00  &  1.00  &  2.5  &  10.4  &  1.9   \\
 Retrograde Orbits & best fit        &  0.572  &  0.99  &  0.52  &  0.5  &   4.6  &  3.5   \\
 Retrograde Orbits & adjusted fit 1  &  0.568  &  1.00  &  0.50  &  0.5  &   4.7  &  3.6   \\
 Retrograde Orbits & adjusted fit 2  &  0.600  &  1.00  &  1.00  &  1.1  &   3.0  &  1.9   \\
\noalign{\smallskip}
\hline
\end{tabular}
\end{table*}

\section*{Acknowledgments}
This work has been supported by the SETI institute (M.~C.), the U.S.
Department of Education under GAANN Grant No. P200A090284 (B.~Q., J.~E.), 
and by the University of Texas at Arlington through its
Research Enhancement Program (M.~C., A.~S.).


\end{document}